\pdfoutput=1
\documentclass[journal=jpcafh,manuscript=article,layout=traditional,letter]{achemso}
\SectionNumbersOn   

\usepackage{graphicx}
\usepackage{booktabs}
\usepackage{microtype}
\usepackage[centertags,intlimits]{amsmath}

\renewcommand{\d}{\mathrm{d}}
\newcommand{\edes}{E^\mathrm{des}}
\newcommand{\edif}{E^\mathrm{diff}}
\newcommand{\ps}[2]{_{\mathrm{#1},#2}}
\newcommand{\s}[1]{_{#1}}
\newcommand{\p}[1]{_\mathrm{#1}}

\title[Interaction of H and H$_2$ with Tholin at Low
Temperatures]{Interaction of Atomic and Molecular Hydrogen with Tholin
  Surfaces at Low Temperatures}

\author{Ling Li}
\author{Hui Zhao}
\author{Gianfranco Vidali}
\affiliation[Syracuse University]{Physics Department, Syracuse
  University, Syracuse, New York 13244}
\author{Yechiel Frank}
\author{Ingo Lohmar}
\affiliation[The Hebrew University]{Racah Institute of Physics, The
  Hebrew University, Jerusalem 91904, Israel}
\author{Hagai B. Perets}
\affiliation[Harvard-Smithsonian Center]{Harvard-Smithsonian Center for
  Astrophysics, Cambridge, Massachusetts 02138}
\author{Ofer Biham}
\email{biham@phys.huji.ac.il}
\affiliation[The Hebrew University]{Racah Institute of Physics, The
  Hebrew University, Jerusalem 91904, Israel}

\date{\today}

\begin{document}

\begin{abstract}
  We study the interaction of atomic and molecular hydrogen with a
  surface of tholin, a man-made polymer considered to be an analogue of
  aerosol particles present in Titan's atmosphere, using thermal
  programmed desorption at low temperatures below $30$~K.  The results
  are fitted and analyzed using a fine-grained rate equation model that
  describes the diffusion, reaction and desorption processes.  We obtain
  the energy barriers for diffusion and desorption of atomic and
  molecular hydrogen.  These barriers are found to be in the range of
  $30$ to $60$~meV, indicating that atom\,/\,molecule-surface
  interactions in this temperature range are dominated by weak
  adsorption forces.  The implications of these results for the
  understanding of the atmospheric chemistry of Titan are discussed.
\end{abstract}

\section{Introduction}

In the last decade, the study of the interaction of hydrogen with
surfaces at low temperatures has become a topic of interest in fields as
different as hydrogen storage~\cite{schlapbach01} and interstellar
chemistry, where molecular hydrogen forms on the surfaces of dust
grains~\cite{combes00, williams07}.  In the latter field, there are
several laboratories studying the mechanisms of reaction of molecular
hydrogen in various space-like environments.  Most laboratory research
on the formation of molecular hydrogen on dust grain analogues, such as
silicates~\cite{pirronello97a, perets07, vidali09}, amorphous
carbon~\cite{pirronello99}, and ices~\cite{roser03, hornekaer03,
  amiaud06}, has shown that this process proceeds by the
Langmuir-Hinshelwood (LH) mechanism~\cite{langmuir18} and is governed by
weak adsorption forces.  In the case of formation of H$_2$ at higher
temperatures ($\approx 300$~K), it was found~\cite{mennella08} that D
atoms sent onto a hydrogen-loaded amorphous carbon surface abstract H
atoms to form HD.  On tholins, an analogue of aerosol particles in
Titan's atmosphere, it was claimed that at high temperatures (above
$150$~K) molecular hydrogen is formed via the Eley-Rideal (ER) abstraction
mechanism~\cite{sekine08a}.

Titan's atmosphere is composed mostly of diatomic nitrogen ($>95\%$) and
methane~\cite{mckay01}.  The dissociation of methane and nitrogen in the
upper atmosphere creates radicals that eventually aggregate in
macroscopic particles that form the well-known brownish haze that
surrounds Titan.  Information on this haze is limited because of the
difficulty of obtaining data from ground observatories or space probes
that would reveal its chemical structure~\cite{szopa06}.  Over the
years, starting with the seminal work of Sagan and Khare~\cite{khare84},
analogues of those particles, called tholins,
have been produced and characterized
in many laboratories~\cite{sagan79, coll99, imanaka04},
and they were found to reproduce the optical
signature of Titan's haze.  Although preparation methods vary, there has
been a convergence about the basic properties of these
analogues~\cite{quirico08}.  They have a general formula C$_x$H$_y$N$_z$
and consist of a disordered chain of highly unsaturated polymers.
Functional groups have been identified\cite{quirico08}.  For our
investigation, we are interested in the addition or removal of hydrogen
via the hydrogenation and abstraction of C$-$C and C$-$N double and triple
bonds~\cite{sekine08a}.

The abundant
presence of unsaturated hydrocarbons is an indirect verification of the
lack of abundant atomic hydrogen in the stratosphere and mesosphere
where a wealth of organics are detected~\cite{lebonnois03}.  To resolve
this discrepancy, it was suggested that hydrogen,
which is produced in the dissociation of CH$_4$,
might recombine to form
molecular hydrogen that then escapes into space~\cite{young84}.  The
formation of molecular hydrogen in Titan's atmosphere follows the same
constraints as the formation of H$_2$ in the interstellar
medium~\cite{duley84}.  The \emph{binary} association of hydrogen atoms
puts the protomolecule in a dissociated state that can make slow
spin-forbidden transitions to the ground state, and the protomolecule
promptly dissociates.  It takes a third particle participating in the
reaction to absorb the excess energy.  In Titan's atmosphere,
however, the density of hydrogen atoms and the total density are
still too small (cf.\
\ref{sec:applications}) to allow a third atom to play this role.  But
formation of H$_2$ taking place \emph{on the surface} of a third body
can be efficient~\cite{bakes03}.  In the case of Titan, the third body
is an aerosol particle.  This view is, however, not universally shared.
For example, a competing mechanism was proposed~\cite{young84}, in which
H$_2$ is catalyzed by C$_4$H$_2$ which is one of the most abundant
hydrocarbon molecules in Titan's atmosphere.  However, it was found that
this scheme is inconsistent with other observations of abundance of
C$_2$- and C$_3$- containing hydrocarbons~\cite{lebonnois03}.

In a recent experiment~\cite{sekine08a}, the formation of HD molecules
was studied by sending D atoms onto tholins.  The desorption of the
reaction product, HD, from the tholin surface was detected by a
quadrupole mass spectrometer, while the change to the surface resulting
from the interaction with D atoms was detected via infrared (IR)
spectrometry in a separate experiment in another apparatus.  The sample
temperature was in a range appropriate for actual aerosol particles,
$160$~K, but experiments were also performed at a higher temperature of
$300$~K.

The formation of HD was attributed to the Eley-Rideal reaction
scheme, sometimes called ``prompt reaction model''~\cite{duley96} in the
astrophysics literature.  In this model, the atom coming from the gas
phase interacts directly with an atom on the surface without first
becoming accommodated to it.  Alternatively, the gas-phase particle
might exchange only part of its energy and move at super-thermal energy
across the surface.  This is called the hot-atom
mechanism~\cite{harris81}.

Most of the surface-catalyzed reactions known in the surface chemistry
literature can be described by the familiar LH model.
The ER reaction or the hot-atom mechanism have been positively
identified only in the 1990's, and mostly on H-plated
single crystal metal surfaces~\cite{rettner94,eilmsteiner96}, H-plated
silicon~\cite{khan07}, H-plated graphite~\cite{zecho02} and H-loaded
amorphous carbon~\cite{mennella08}.  There are two major signatures to
look for in order to identify the ER reaction model: the detection of
super-thermal energy in the HD leaving the surface, and the time
dependence of the HD yield during irradiation of the surface with D
atoms.  Such irradiation depletes the surface of H atoms by their
reaction with incoming D atoms, and hence the measured HD yield
decreases exponentially with the irradiation time.

When the tholin sample was first exposed to D atoms, a small increase in
the HD signal was measured~\cite{sekine08a}, but there was little change
in the HD signal over time.  A similar observation was made for the
formation of HD from the interaction of D with hydrogen-loaded amorphous
carbon at room temperature~\cite{mennella08}.  In this experiment, the
cross section for the reaction between an adsorbed H atom and an
incoming D atom was obtained from IR data, and was found to be almost
two orders of magnitude smaller than the one measured for the
interaction of energetic H atoms on a graphite surface~\cite{zecho02}.
The weak time dependence of the HD signal was attributed to this very
small value of the cross section~\cite{mennella08}.  Alternatively, we
could interpret this result as caused by a small probability of reaction
when a D atom hits an H atom that is on the surface.

In this paper we report on a study of the formation of molecular
hydrogen (specifically, HD and D$_2$) on tholins at lower temperatures,
below $30$~K.  At these low temperatures the diffusion and desorption
processes are slower and the residence times of the weakly adsorbed
atoms and molecules on the surface are longer.  This leads to reduced
noise levels and enables us to determine whether the LH or the
abstraction mechanism are operative.  We obtain activation energies for
the diffusion and desorption of hydrogen atoms (H and D without
distinction) and molecules (HD and D$_2$) on or from the surface.  To
this end, we use thermal desorption spectroscopy coupled with an
analysis using rate equation models.  This work builds on methodologies
developed in previous studies of the formation of molecular hydrogen on
analogues of interstellar dust particles, such as
silicates~\cite{perets07, vidali07, vidali09}, carbonaceous
materials~\cite{katz99}, and ices~\cite{manico01, roser03, perets05}.

The Paper is organized as follows.  In \ref{sec:experiment} we describe
the experimental setup and the type of measurements performed.
\ref{sec:results} presents the results of these measurements.  We then
explain in detail the rate equation model (\ref{sec:rate-eq}) that we
employ.  The experimental data is analyzed using this model in
\ref{sec:analysis}, which contains our main results on the energy
landscape.  We discuss applications of our results in
\ref{sec:applications}, before concluding with our Summary
(\ref{sec:summary}).

\section{Experimental Methods}
\label{sec:experiment}

\subsection{Apparatus}

The apparatus used for these experiments is the same as the one employed
to study the formation of molecular hydrogen on interstellar dust grain
analogues~\cite{vidali05}.  It consists of two atom beam lines and a
sample\,/\,detector chamber.  In the beam lines, hydrogen and deuterium
gas is dissociated by two radio frequency sources.  The beam in each
line is formed in three differentially pumped stages and is highly
collimated.  A metrological laser coaxial with the line is used to align
the beam such that the two beams strike the same spot on the sample.
The partial pressure in the third stage is in the $10^{-8}$~Torr range,
and this stage is separated from the ultra-high vacuum chamber by a
$2.5$~mm collimator.  The main chamber has a base pressure in the low to
mid $10^{-10}$~Torr range.  The detector, a high performance Hiden
triple pass quadrupole mass spectrometer, is mounted on a rotatable
flange, and is configured so it can measure both the flux coming from
the sample and the one from the beams; different masses can be probed
simultaneously.  The sample is mounted on a cold finger, surrounded by a
cold copper shield to improve cooling.  In the experiments reported
here, the cold finger was cooled with liquid helium and the desired
sample temperature was obtained using a cartridge heater (Lakeshore).
The temperature was measured with a calibrated silicon diode
(Lakeshore).

We use beam lines since they allow to control the kinematic conditions
of irradiation of the sample to very good extent.  Through differential
pumping and the use of a mechanical chopper, low doses of H and D can be
sent.  Although the flux is understandably orders of magnitude higher
than in actual space conditions, it is much less than traditionally
achieved in laboratory experiments
(cf.~\ref{sec:findflux,sec:applications}).  The reason for having two
lines is that even if we use a very effective dissociation source
($\approx 75\%$), a small fraction of D$_2$ is transmitted on towards
the sample.  Deuterium is used instead of hydrogen because it is easier
to detect HD as the product of reaction than H$_2$, which is the main
residual gas in a well-baked ultra-high vacuum stainless steel chamber.

Additionally, the apparatus was instrumented with an IR
spectrometer (Nicolet FT-IR 6300)
in the reflection-absorption infrared spectroscopy (RAIRS) configuration
in order to study in-situ surface modifications.
The
IR light from the source is sent into the apparatus via a
differentially pumped MgF window. The beam strikes the sample at
glancing incidence ($\approx 84^{\circ}$) and is then collected by a
liquid nitrogen cooled mercury cadmium telluride
detector placed outside the vacuum chamber.  This
arrangement grants high sensitivity and allows to perform the IR
measurements without breaking the vacuum.

\begin{figure}
  \centering
  \includegraphics{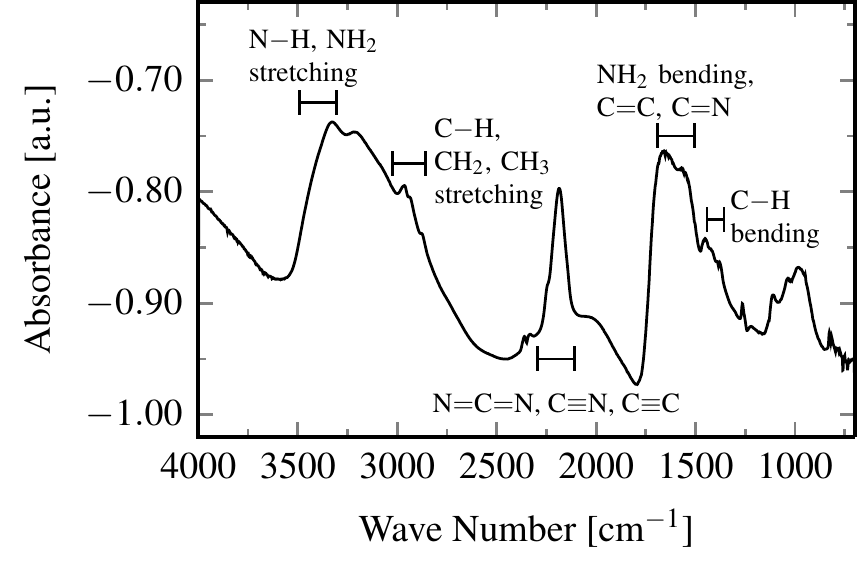}
  \caption{RAIRS spectrum of the tholin sample prior to exposure to
    hydrogen and deuterium atoms.  Characteristic bands are shown.}
  \label{fig:FTIR}
\end{figure}

The sample was prepared by Prof.\ Mark Smith's laboratory (University of
Arizona).  The tholin films were deposited on a gold plated copper disc
of $1/2$~inch diameter and $1$~mm thickness.  The films are thick ($>
500$~nm) and were produced over 4 days in a $10$~kV AC discharge in
$2\%$ Methane and $98\%$ N$_2$ at $14$~Torr and at $300$~K.
Preparation and characterization methods of tholins made by arc
discharge---as the ones used here---have been described
before~\cite{neish10,neish09}.  This preparation method yields tholins
that are similar but not identical to the ones prepared at lower
pressure.  However, the higher pressure is necessary to obtain a film
thick enough for use in these experiments.  More details and comparisons
can be found in the recent literature~\cite{quirico08, neish09, neish10,
  szopa06}.  Characterization using Fourier-transform ion cyclotron mass
spectrometry is available as well~\cite{sarker03}.  A Fourier transform
infrared (FTIR) spectrum of
the sample prior to exposure to hydrogen is shown in \ref{fig:FTIR}.

\begin{figure}
  \centering
  \includegraphics[width=8.25cm]{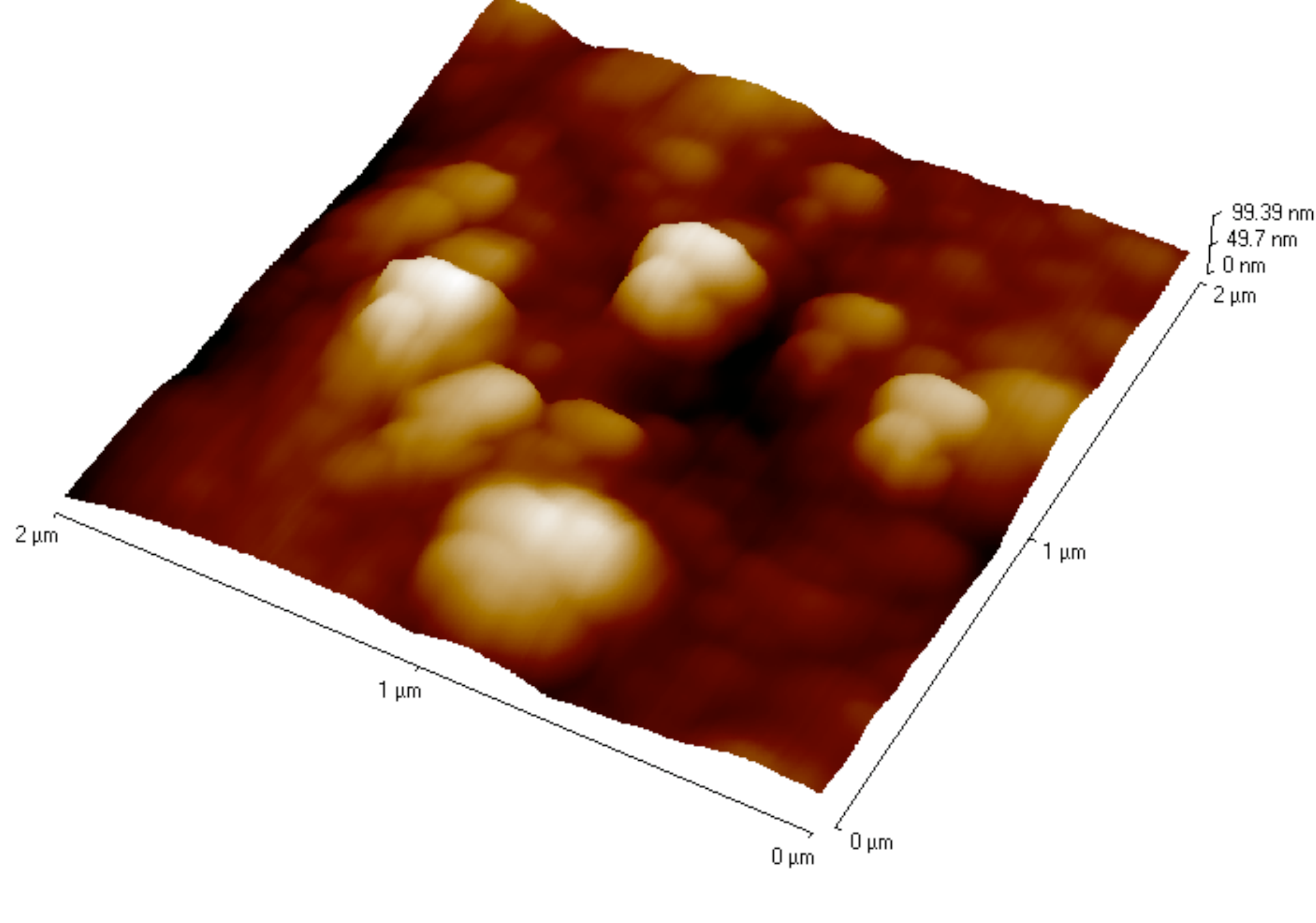} 
  \caption{Three-dimensional rendering of an AFM scan on a tholin
    sample.  The area covered is $4$~$\mu$m$^2$ and the maximum height
    variation over this area is $100$~nm.}
  \label{fig:example}
\end{figure}

Each sample was
sent in a sealed pouch and was mounted on the sample holder while
working under a flow of dry nitrogen.  The vacuum chamber was
pressurized with dry nitrogen gas before the insertion of the sample
holder.  Several samples were used in the experiments; for cleaning
purposes, while in vacuum, the samples were taken to $350$~K.
One of them was also analyzed using an Atomic Force Microscope
(AFM) (KLA-Tecnor P16+).
\ref{fig:example} shows that on a $2$~$\mu$m linear scale, there
is a height-to-height variation of $100$~nm.

\subsection{Experimental Procedures}
\label{sec:findflux}

\begin{figure}
  \centering
  \includegraphics{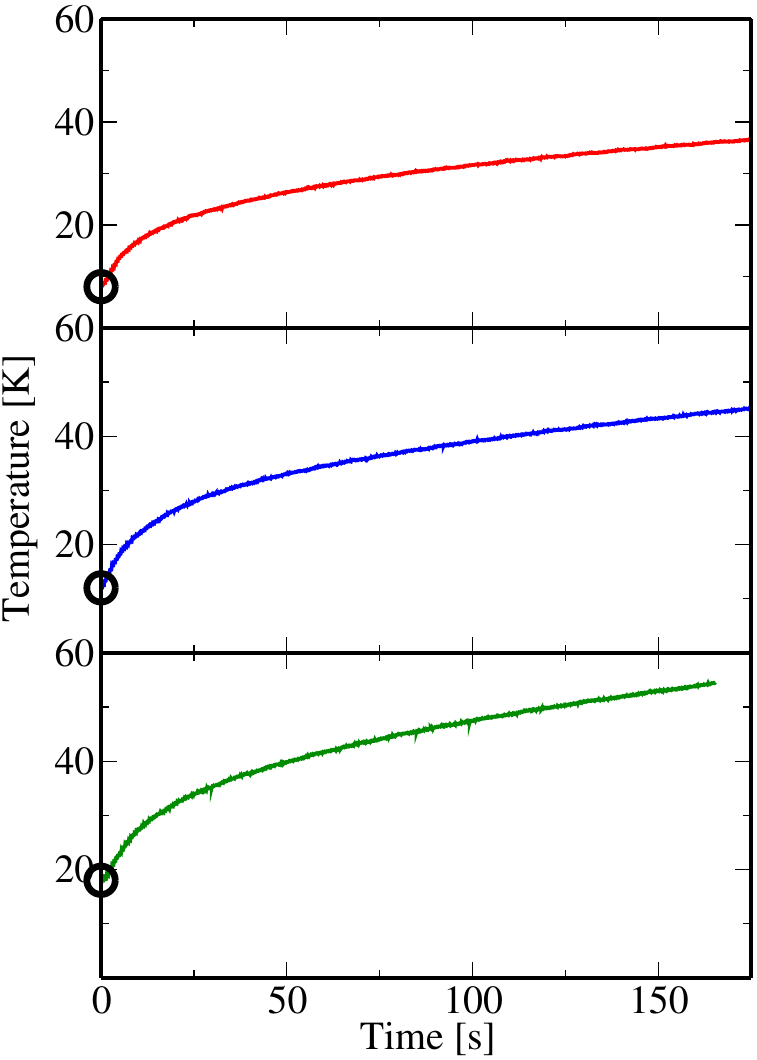}
  \caption{Temperature ramp of the sample for different irradiation
    temperatures: $8$~K (top), $12$~K (middle) and $18$~K (bottom).
    These curves as measured in the experiment with irradiation of HD
    molecules.  The irradiation temperature is marked by the circle on
    the temperature axis.}
  \label{fig:T-t-for-HD-mol}
\end{figure}

In temperature programmed desorption (TPD) experiments, the surface is
first exposed to beams of atoms or molecules, at a fixed sample
temperature and for a set amount of time.  The sample
temperature is then ramped up and the products desorbing from the
surface are detected in real time.  In the experiments presented here,
we start the heating at a rate of about $6$~K$/$s that decreases to
$0.06$~K$/$s eventually.
During the experiment the amount $\d N$ of gas particles detected within
a small (and constant) time interval $\d t$ is recorded.
Simultaneously, we measure the sample temperature $T(t)$ (see
\ref{fig:T-t-for-HD-mol} for an example).  To be able to compare several
desorption measurements, we need to convert the rate $\d N/\d t$ to the
detection ``rate'' $\d N/\d T$ with respect to temperature $T$.
In practice, we first fit a function $T(t)=\mathrm{const}\cdot t^\gamma$
to the temperature ramp ($t=0$ corresponds to the start of heating).
From this we determine $\d T/\d t$, unaffected by noise in the original
measurement.  We then obtain
\begin{equation}
  \frac{\d N}{\d T}(T)
  =\frac{\d t}{\d T} \cdot \frac{\d N}{\d t}(t).
\end{equation}
In the rate equation simulation we use a similar procedure.  After every
time step of the Runge-Kutta procedure we calculate the temperature
change during that step, according to the fit for $T(t)$.  From this we
directly obtain the value of $\d N/\d T$ in the simulation.

The shape and the position of peaks in $\d N(T)/\d T$ provide
information on the kinetics and energetics of the reactions, as
illustrated below.  In the experiment performed using
simultaneous H and D beams, one probes the amount of HD formed on the
surface.  HD can form on the surface either rapidly (compared to
laboratory time scales) due to fast diffusion, essentially while the
sample is still being irradiated with H and D, or it can form during the
heat pulse when the H and D atoms that became adsorbed on the surface
during the irradiation phase become mobile, encounter each other and
form HD.

The fluxes of the beams are measured using the rotatable quadrupole
mass spectrometer without exposing the samples to the beams.
The detector is placed between the beam lines (which are $38^\circ$
apart) and the 
signals are recorded in real time.
The measured effective beam density
reaching the surface (after chopping) is 
$f_\mathrm b=5\times10^{10}$~cm$^{-2}$~s$^{-1}$.  
The effective flux to the surface is estimated as follows.  We assume a
density $s\approx5\times10^{14}$~sites$/$cm$^2$ of adsorption sites on
the tholin surface.  This is a reasonable value based on data for other
materials such as silicates~\cite{perets07}.  
We thus obtain an effective flux
of $f =f_\mathrm b/s =1\times10^{-4}$~monolayers~(ML)$/$s.

\section{Experimental Results}
\label{sec:results}

\subsection{Eley-Rideal Prompt Reaction}
In the ``prompt reaction'' scheme, a D atom abstracts a hydrogen atom on
the surface, forming an HD molecule which leaves the surface.  Such
reaction has been observed in H-plated metals and graphite and on
H-loaded amorphous carbon.  The typical cross section is expected to be
of the order of a few~\AA$^2$~\cite{eilmsteiner96, khan07, zecho02}.
However, as already mentioned in the
Introduction, a much smaller value of $\sim 3 \times10^{-2}$~\AA$^2$ was
found on H-loaded amorphous carbon~\cite{mennella08}.
During the irradiation phase, the detector is
positioned to measure any HD coming off the surface.  For irradiation of
H and D on tholin samples at low temperature, we find this contribution
indistinguishable from the background.  We thus conclude that the prompt
reaction mechanism is inefficient under the physical conditions used
here.

\subsection{Langmuir-Hinshelwood Reaction}
\begin{figure}[htbp]
  \centering
  \includegraphics{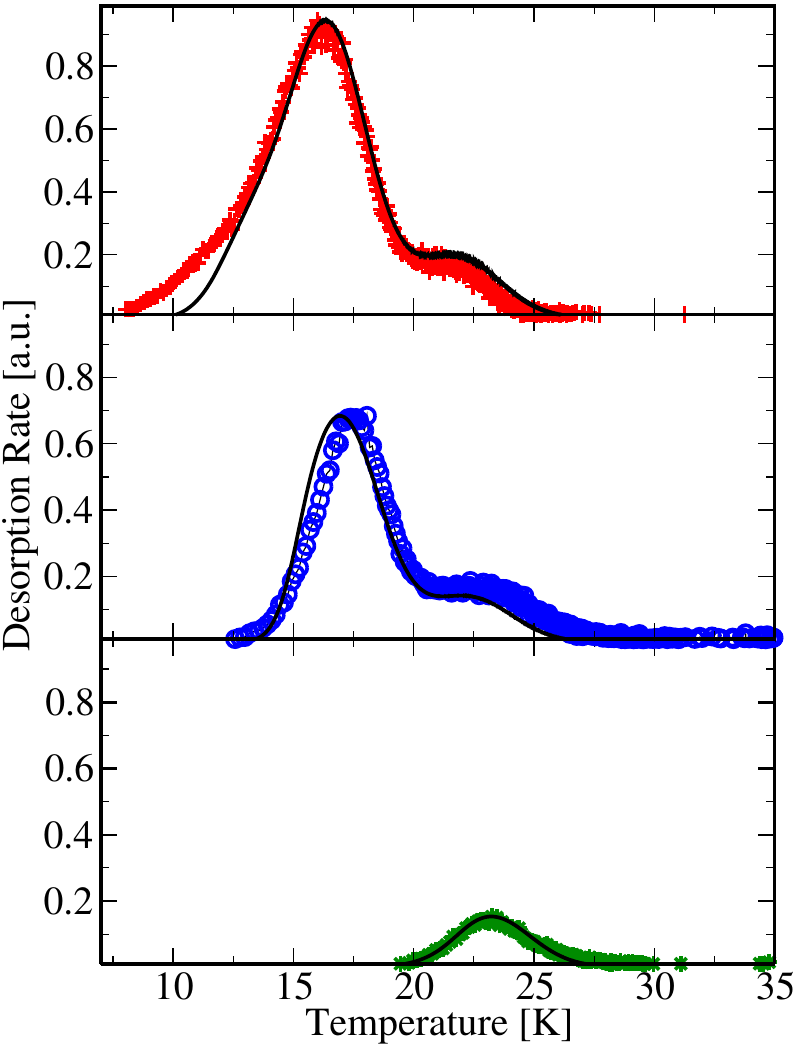}
  \caption{Desorption rate of HD molecules vs.\ temperature after
    irradiation of the surface with HD molecules.  Irradiation
    temperatures $8$~K (top), $12$~K (middle) and $18$~K (bottom).
    Colored symbols: Experimental data.  Black lines: Fit using the rate
    equations (reduced model).}
  \label{fig:HD-mol}
\end{figure}

These experiments are similar in  methodology to the ones we carried out
on  interstellar dust  grain  analogues~\cite{vidali05, vidali09}.
In \ref{fig:HD-mol} we present the desorption rate of HD molecules vs.\
temperature, after irradiation of HD molecules on the surface at surface
temperatures of $8$~K, $12$~K and $18$~K.
The irradiation time was $120$~s at $8$~K, while for the higher
temperatures the sample was irradiated for $240$~s.
The peak positions of a trace can be related to the activation
energy of desorption~\cite{yates85}.
The trace obtained after irradiation at $8$~K consists of two peaks---a
large peak at low temperature and a small peak at a somewhat higher
temperature.
This indicates that there are at least two types of adsorption sites for
HD molecules on the tholin surface.
The relative areas below the two peaks suggest that there is a large
number of shallow binding sites and a much smaller number of deep
binding sites.
The trace obtained after irradiation at $12$~K exhibits a similar shape;
however, its peak heights are decreased and its low-temperature edge is
shifted to higher temperatures.
This can be explained by the fact that at a surface temperature of
$12$~K those molecules adsorbed in shallow sites may quickly desorb
already during irradiation.
This effect is even more pronounced in the case of irradiation at
$18$~K, where the low-temperature peak has completely vanished.

\begin{figure}[htbp]
  \centering
  \includegraphics{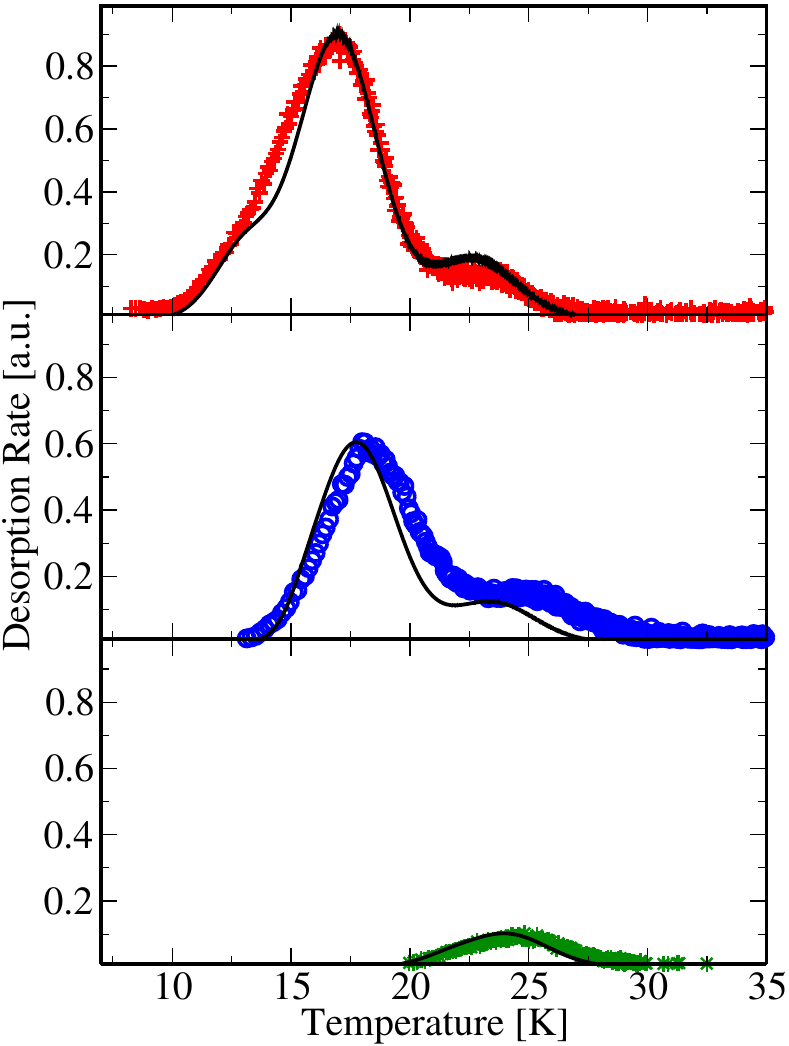}
  \caption{Desorption rate of D$_2$ molecules vs.\ temperature after
    irradiation of the surface with D$_2$ molecules.  Irradiation
    temperatures and symbols as in \ref{fig:HD-mol}.}
  \label{fig:D2-mol}
\end{figure}

A similar set of TPD traces for D$_2$ molecules is shown in
\ref{fig:D2-mol}.  Irradiation temperatures and times are as for the HD
case.  Everything we said about the traces for HD molecules applies here
as well, but the peaks of traces of D$_2$ desorption are shifted to
higher temperature with respect to the peaks of HD.  This is due to the
isotope effect.  If we take the trapping potential to be a harmonic
oscillator with an (unknown) ``spring'' constant $k$, then the lowest
energy level for a molecule of mass $m$ in this potential is
$\hbar\sqrt{k/m}/2$.  The higher atomic mass of D$_2$ molecules
therefore leads to a lower ground state energy, which in turn leads to a
larger activation energy for desorption.

As previously found in the analysis of HD and D$_2$ formation on
amorphous silicates at different temperatures~\cite{vidali09}, there is
a common trailing edge in the three traces.  This corroborates our
fundamental assumption that the frequency and magnitude of certain
energy barriers are properties of the surface
morphology, and as such do not depend on the surface temperature during
irradiation.  An analysis of the shapes of these traces provides
information on the distribution of binding energies, as we will show
below.

\begin{figure}  
  \centering
  \includegraphics{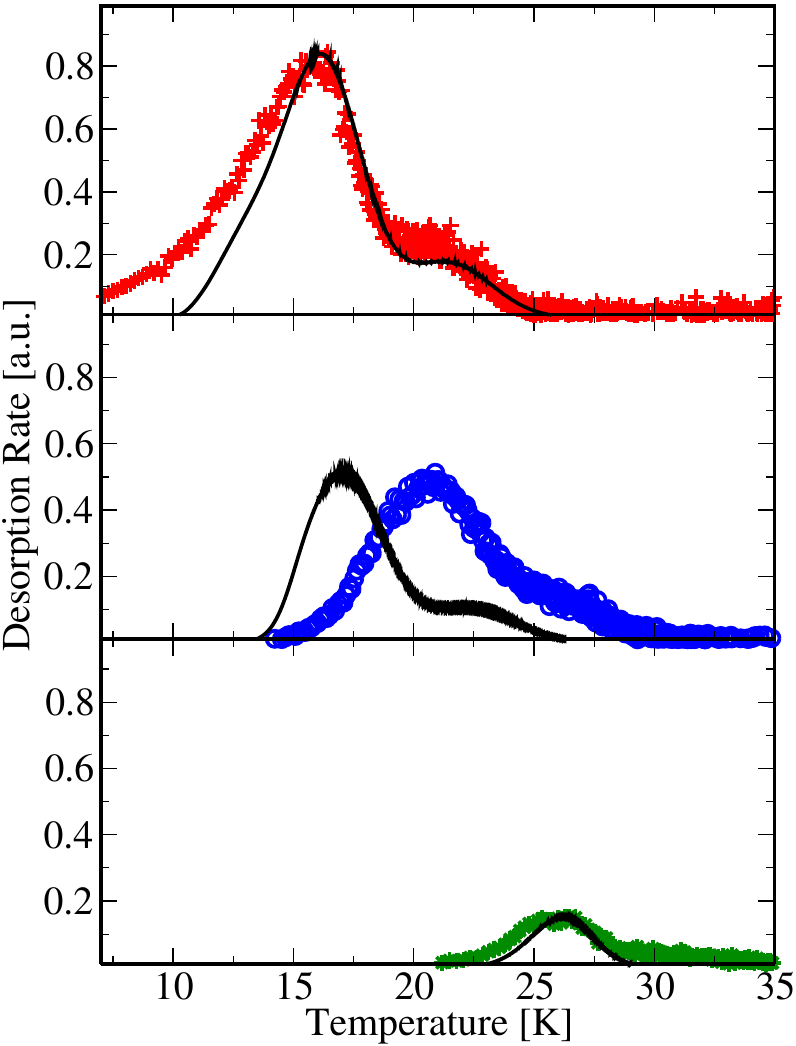}
  \caption{Desorption rate of HD molecules vs.\ sample temperature after
    irradiation of the surface with H and D atoms.  Irradiation
    temperatures $8$~K (top), $12$~K (middle) and $22$~K (bottom).
    Colored symbols: Experimental data.  Black lines: Fit using the rate
    equations (reduced model).}
  \label{fig:H-D-at}
\end{figure}

\begin{figure}  
  \centering
  \includegraphics{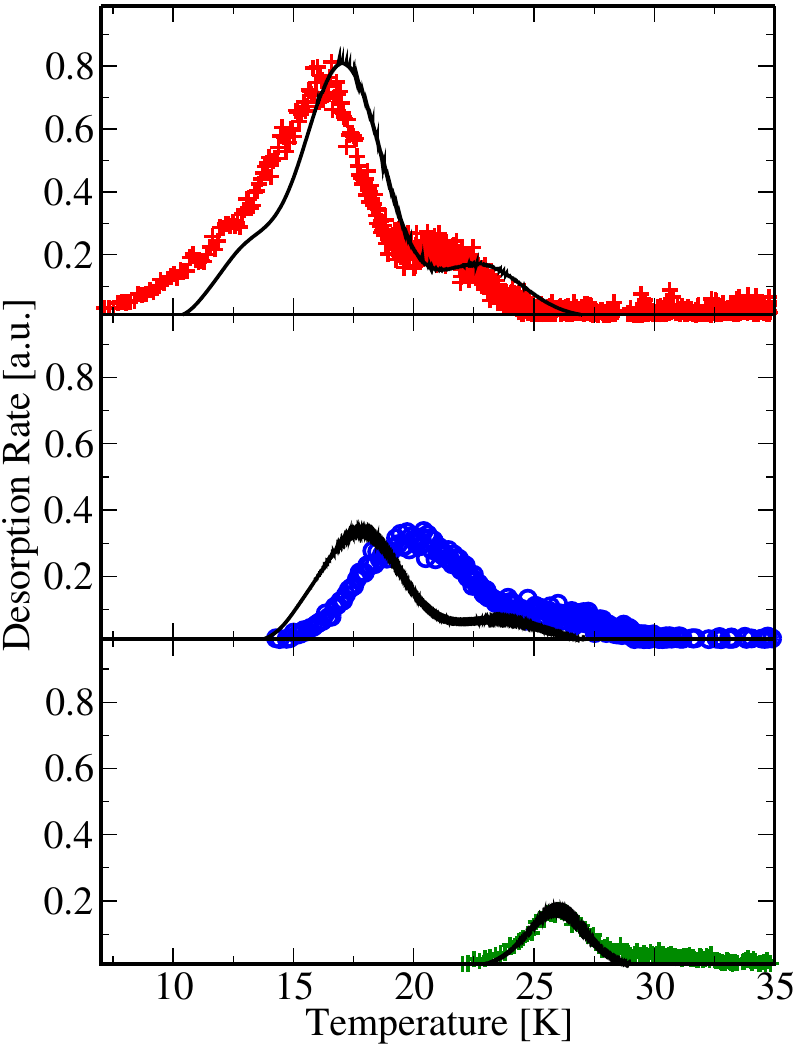}
  \caption{Desorption rate of D$_2$ molecules vs.\ sample temperature
    after irradiation of the surface with D atoms.  Irradiation
    temperatures and symbols as in \ref{fig:H-D-at}.}
  \label{fig:D-at}
\end{figure}

\begin{figure}  
  \centering
  \includegraphics{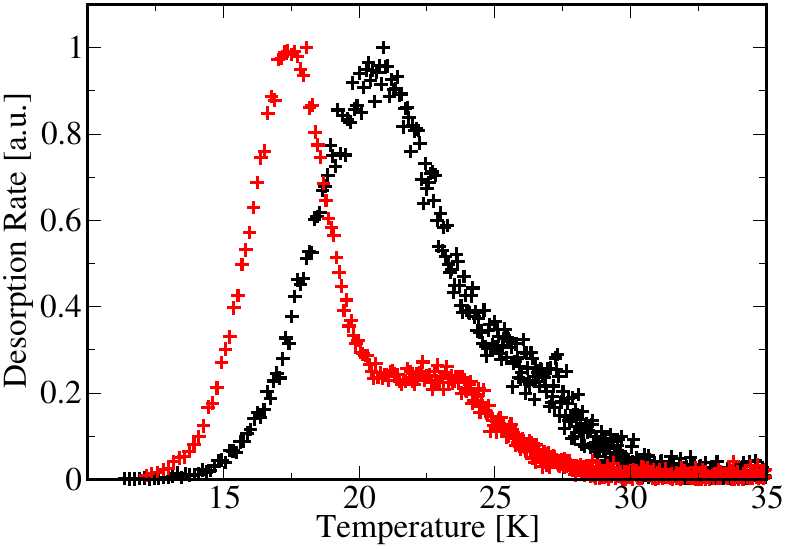}
  \caption{Comparison of the desorption rate of HD molecules vs.\ sample
    temperature, after irradiation of the surface with HD
    molecules (red pluses) and after irradiation with H and D atoms
    (black pluses), both at $12$~K.  The traces are normalized.}
  \label{fig:HD-mol-vs-atoms}
\end{figure}

\ref{fig:H-D-at,fig:D-at} show TPD curves after irradiation with H and D
\emph{atoms} and detecting HD, and with only D \emph{atoms} and
detecting D$_2$, respectively.
The irradiation time is $120$~s throughout, and
irradiation temperatures are $8$~K, $12$~K and $22$~K.  In both cases,
for HD as well as for D$_2$ detection, we observe that the
traces have shapes very similar to their corresponding counterpart
obtained from irradiation with molecules (\ref{fig:HD-mol,fig:D2-mol}).
Depending on the irradiation temperature, however, peaks of the TPD
curves obtained by the association of atoms are shifted to higher
temperatures relative to the ones obtained after irradiation with
molecules.  We observe this shift for both HD and D$_2$ formation for
the intermediate irradiation temperature of $12$~K.  We cannot explain
this feature easily, and will return to this in our analysis in
\ref{sec:analysis}.
For the case of HD, a comparison between molecule and atomic irradiation
is shown in \ref{fig:HD-mol-vs-atoms}.
Note that the shift between the bottom panels of
\ref{fig:HD-mol,fig:H-D-at}, and of \ref{fig:D2-mol,fig:D-at},
respectively, first and foremost reflects the fact that atomic
irradiation was done at $22$~K, while molecular irradiation was done at
$18$~K.  Performing both types of experiment after irradiation at
$18$~K, we still found a shift (not shown), but of significantly smaller
size ($\approx 1$~K) than for the comparison at $12$~K.

The diminishing intensity of the signal with increasing irradiation
temperature indicates that the processes involved in the formation of HD
from H and D atoms are governed by weak physical adsorption forces,
implying lower sticking probability of H and D at higher sample
temperature.  The widths of the traces are larger than the ones expected
if there were only a single activation energy for desorption.  Rather,
similarly to what we detected in the formation of HD on amorphous
silicates~\cite{vidali09} but to a lesser degree, there is a range of
activation energies.

We further note that the trailing edges of traces obtained after
irradiation at different temperatures do \emph{not} coincide for atomic
irradiation, in contrast to what we found for molecular irradiation.
This implies that the process of molecular formation (whether during
irradiation or during the subsequent heating) affects the distribution of
molecules to the different types of adsorption sites, in a way that
depends on the surface temperature.

\section{Rate Equation Model}
\label{sec:rate-eq}

In all TPD curves considered here we observe that most of the hydrogen
is desorbed well before $40$~K.  We therefore conclude that the
particles are trapped in physisorption potentials and are only weakly
adsorbed.  We also assume that the mechanism of formation of H$_{2}$ (or
HD or D$_{2}$) is the LH scheme, as there is no evidence of prompt
reaction.

The analysis of TPD experiments usually starts with the Polanyi-Wigner
expression for the desorption rate,
\begin{equation}\label{polanyi-wigner}
  R(t)=\nu {N(t)}^{\beta} \exp\left(\frac{-E}{k_\mathrm B T}\right).
\end{equation}
In this expression, $N(t)$ is the total number of atoms on the surface,
$\beta$ is the order of desorption, and $\nu$ is the vibration frequency
of the particle in the potential well where it is bound, also referred
to as the attempt frequency.
The effective activation energy is denoted by $E$,
and $T=T(t)$ is the time-dependent
surface temperature.
One important assumption we make here is that all the surface
properties, such as the attempt frequency and the energy barriers, are
independent of temperature and population.  This assumption is justified
due to the low coverage of the surfaces during a TPD experiment
($<0.01$~ML), but it might be violated at high coverages~\cite{barton78}.
Analyses of TPD experiments using rate equations have been reported
previously~\cite{katz99, cazaux04, perets05}. 
Here we introduce two models for describing the TPD
experiments (for molecule as well as for atomic irradiation), a complete
model accounting for all possible processes in the system, and a reduced
one.  We show that the reduced model gives good results in fitting the
experimental data.

\subsection{Complete Model for Irradiation with Molecules}
\label{sec:compl-model-mols}
We first introduce the model for molecules only.  It will be
modified below to deal with atoms and their reactions.
Molecules of a given species are sent onto the surface, and if they
impinge onto an empty adsorption site, they stick to it with a certain
probability.  The rate at which particles stick to the (empty) surface is
the \emph{effective} flux $f$ (in ML$/$s).  Afterwards they
may hop from a site to any of the neighboring sites, and may also
desorb.
Throughout we assume the rates for both processes to be
thermally activated, with a common fixed attempt frequency $\nu$ (taken
standardly to be $10^{12}$ s$^{-1}$).  The hopping rate is given by
\begin{equation}
  D = \nu \exp\left(\frac{-\edif}{k_{\mathrm B}T}\right),
\end{equation}
where $\edif$ is the activation energy for diffusion, and $T(t)$ is the
time-dependent temperature of the surface during the TPD experiment.
The activation energy in general depends on the particle species and the
type of site it is located at, both of which will later appear as
indices on $\edif$ as well as on $D$.
Similarly, the desorption rate reads
\begin{equation}
  W = \nu \exp\left(\frac{-\edes}{k_{\mathrm B}T}\right),
\end{equation}
where $\edes$ is the activation energy for desorption, which can
also depend on the site type and the species.
As the surface temperature increases, both the hopping rate and the
desorption rate rapidly increase.

We now make several assumptions.
First, we assume a given density of
adsorption sites on the surface, each containing at most one molecule.
We recognize that for the diffusion and desorption of H$_{2}$ (HD and
D$_2$) molecules the surface is not homogeneous.  We model this by using
several
types of sites, distinguished by \emph{Greek} indices in our notation.
Each type $\alpha$ has its own \emph{average} energy barrier for
diffusion, $\edif_{\alpha}$, and for desorption, $\edes_{\alpha}$.
Additionally, the energies for specific sites of type $\alpha$ are
distributed
according to a normal distribution around the average energy.  The
standard deviation of this normal distribution is labeled
$\sigma_{\alpha}$.
In order to retain detailed balance we set $\edes_{\alpha}
=\edif_{\alpha} +\Delta E$, where $\Delta E$ is an overall constant.  An
analogous relation holds for the desorption and diffusion barrier of
each individual site, hence both energy distributions have the same
standard deviation $\sigma_{\alpha}$.
Each type $\alpha$ of sites constitutes a fraction $\rho_{\alpha}$ of the entire
surface such that $\sum_{\alpha}\rho_{\alpha}=1$.

\begin{figure}
  \centering
  \includegraphics{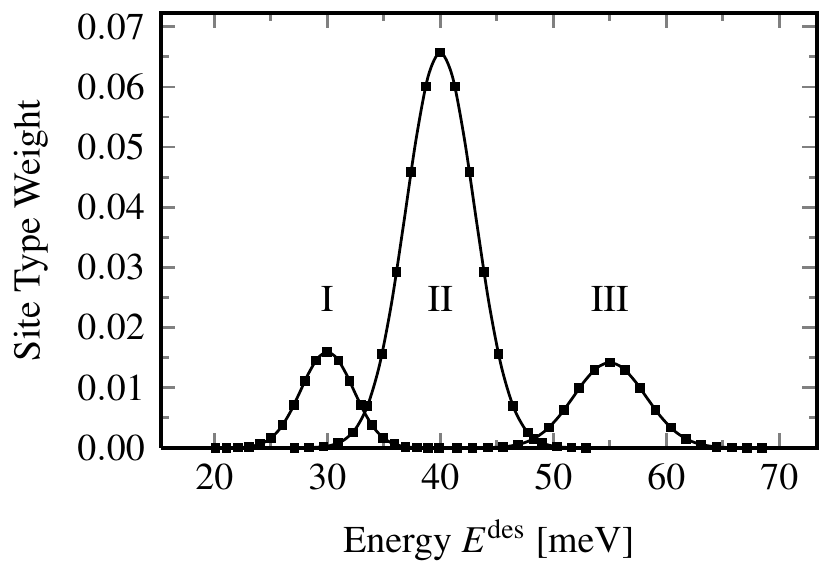}
  \caption{A schematic picture of the sampling of energies to model the
    normal distribution around mean energies, for three types of sites.
    The three normal distributions of binding energies for site types I,
    II, and III, respectively (solid lines), and the 21 sample energies
    used to model each distribution (black squares) are shown.
    Mean values and
    widths are taken from the results we obtain for HD molecules (see
    \ref{sec:analysis}).}
  \label{fig:site-samples}
\end{figure}

In a rate equation model, we cannot employ a continuous distribution of
energies, and neither can we model a particular realization of surface
site energies.  (Note that a continuous distribution of binding energies
can be obtained by direct inversion of TPD traces~\cite{amiaud06,
  barrie08, vidali10}, but this method does not include the possibility
of simultaneous recombination processes.)  We approximate the
distribution of binding energies of a certain site type $\alpha$ around
the mean value by using $21$ different binding energies as samples for
each type.  The energy values are equidistantly spaced between
$\edes_\alpha-3\sigma_\alpha$ and $\edes_\alpha+3\sigma_\alpha$.  The
fraction of sites at any given sample energy is chosen proportional to
the value of the normal probability distribution function with mean
$\edes_\alpha$ and standard deviation $\sigma_\alpha$ at that energy.
This sampling of energies is schematically depicted in
\ref{fig:site-samples}.
The individual sample will be denoted by a Latin index, e.g.,
$\edes\s{i}$, and the weights of sites of the $21$ sample energies add up
to the total fraction of the `fundamental' site type, $\rho_\alpha =
\sum_{i \in \mathrm{type}\; \alpha} \rho\s{i}$, hence $\sum_i\rho\s{i} = 1$
again.  For the rate equation model, we effectively just have a large
number of different site types, and we do not need to distinguish what
basic type of site $\alpha$ they belong to.  Hence, we shall still speak
of different `types' of sites also for the different $i$.

Let $n$ be the coverage of molecules on the surface (in ML)
and $n\s{j}$ that part of the total surface coverage which is trapped in
sites of type $j$.  Then $n\s{j}\leq\rho\s{j}$ and $n=\sum_{j}n\s{j}
\leq1$.
The set of rate equations for our model reads
\begin{equation}
  \label{ratefullmol}
  \begin{split}
    \frac{\d n\s{i}}{\d t}
    &=f(\rho\s{i}-n\s{i})
    +\sum_{j\ne i} D\s{j} n\s{j} (\rho\s{i}-n\s{i})\\
    &\quad-D\s{i} n\s{i} \sum_{j\ne i}(\rho\s{j}-n\s{j})
    -W\s{i} n\s{i}.
  \end{split}
\end{equation}
The first term on the right hand side of \ref{ratefullmol} covers the
incoming flux of molecules.  The value of the effective flux $f$ is
found as described in \ref{sec:findflux}.  This term
also accounts for LH rejection, such that all molecules impinging on top
of a site already occupied are rejected.
The second term describes the diffusion of molecules, arriving
\emph{from} other sites $j$.  Likewise, the third term is the loss of
molecules by diffusion from sites of type $i$ to sites of any other type
$j$.  The hopping rate $D\s{j}$ on sites of type $j$ is
determined by the activation energy for diffusion on such sites,
$\edif\s{j}$.
The last term models desorption of molecules, with a desorption rate
$W\s{i}$ determined by $\edes\s{i}$, the activation energy for
desorption from a site of type $i$.  The rate at which molecules are
detected during the experiment is therefore proportional to the total
desorption rate
\begin{equation}
  R = \sum_{i} W\s{i} n\s{i}.
\end{equation}
This model accounts for all possible processes of motion of molecules:
molecules can diffuse from any site type to any other site type and they
can desorb.

To obtain the energy barriers for molecule diffusion and desorption, we
examined the TPD curves of experiments in which molecular HD or D$_{2}$
were deposited on the surface during the irradiation phase and were
later desorbed from the surface as the temperature increased (see
\ref{fig:HD-mol,fig:D2-mol}).
These curves show a broad distribution of temperatures at which
molecules desorb from the surface, and two or three desorption peaks
within this range.  We therefore use three average activation energy
barriers, each with an additional normal distribution of the energy
(totaling $63$ sample energies $E\s{i}$).
The fraction $\rho_\alpha$ (of sites of type $\alpha$) found here is
kept constant for all fits and experiments, see \ref{sec:analysis} for
the results.

\subsection{Complete Model for Irradiation with Atoms}
\label{sec:compl-model-atoms}

We will now modify the model to describe the following situation.  H
atoms (precisely, H and D, or only D) are sent onto the surface and
stick to empty sites with a certain effective flux.  Atoms explore the
surface just like molecules, and they can desorb as well.  Additionally,
when two atoms meet, they form a molecule.

To keep the number of parameters small, we do not distinguish between H
and D atoms in our model, both species implicitly sharing the activation
energies for desorption and diffusion.  Experimentally, however, the
isotope effect is observed and yields different energies for H$+$D vs.\
D$+$D experiments.  The implications of this approximation have been
examined~\cite{vidali07}.  The activation energies of desorption and
diffusion of D atoms were raised by about $10\%$ (this is comparable to
the isotope effect measured in H and D scattering experiments from a
graphite surface~\cite{ghio80}).  The TPD traces hardly changed because
they mainly depend on the energetics of the most mobile species.

All assumptions detailed in the molecule model will remain in effect,
including the uniform standard attempt frequency.
Each adsorption site can now hold either an H or a D atom, or an HD or
D$_2$ molecule.  To simplify equations we assume that molecules and
atoms on the surface do not encounter each other while hopping; this is
justified due to the low coverage during TPD experiments.
In contrast to the situation for molecules, we further assume that all
sites are identical both for hopping and desorbing of atoms, hence the
energy barriers $\edes\p{H}$ and $\edif\p{H}$ are uniform all over the
surface.  Here and in the following, all quantities will get an extra
index for the atom or molecule species.  However, by our above comment,
we do not distinguish between H and D, so that all atoms are labeled H,
and all molecules H$_2$.

We denote by $n\p{H}$ the coverage of H atoms on the surface (in ML) and
by $n\ps{H_2}{j}$ that part of the total surface coverage of H$_2$
molecules (in ML) which is trapped in sites of type $j$.  Consequently,
$n\ps{H_2}{j} \leq \rho\s{j}$ and $n\p{H} \leq 1$.
The new set of rate equations is now given by
\begin{subequations}\label{ratefull}
  \begin{align}
    \label{ratefulla}
    \frac{\d n\p{H}}{\d t}
    &=f(1-n\p{H}) -W\p{H} n\p{H}
    -2D\p{H}n\p{H}^2,\\
    \label{ratefullb}
    \begin{split}
    \frac{\d n\ps{H_2}{i}}{\d t}
    &=\frac{\rho\s{i}-n\ps{H_2}{i}}{\sum_{j} (\rho\s{j}-n\ps{H_2}{j}) }
    D\p{H} n\p{H}^2\\
    &\quad+\sum_{j\ne i} D\ps{H_2}{j} n\ps{H_2}{j} (\rho\s{i}-n\ps{H_2}{i})\\
    &\quad-D\ps{H_2}{i} n\ps{H_2}{i} \sum_{j\ne i} (\rho\s{j}-n\ps{H_2}{j})
    -W\ps{H_2}{i} n\ps{H_2}{i}.
  \end{split}
\end{align}
\end{subequations}
The first term on the right hand side of \ref{ratefulla} describes the
incoming flux of \emph{atoms}, including LH rejection.  $f$ is the
\emph{effective} flux as explained before, only for atoms.
The second term describes the desorption of atoms.  The desorption
rate $W\p{H}$ is governed by $\edes\p{H}$ as specified above.
The last term in \ref{ratefulla} is the recombination term, i.e., the
rate of molecule formation on the surface.  Here the hopping rate
$D\p{H}$ is determined by $\edif\p{H}$.
\ref{ratefullb} is the set of equations describing the dynamics of the
molecules on the surface.
The first term in \ref{ratefullb} consists of two factors.  The second
factor, $D\p{H}n\p{H}^2$ is the rate of molecule production as given by
\ref{ratefulla}; we assume that all molecules remain on the surface once
they are formed, as observed experimentally.  The first factor of this
term distributes the molecules between the different types of sites
according to the distribution of \emph{free} sites among them.
The remaining terms of \ref{ratefullb} have been explained above in the
molecular model.  The only difference is in the notation, since all
quantities now carry the additional species label.

\subsection{Reduced Models}

The complete model for irradiation with molecules successfully fits the
TPD experiments (not shown),
but the best fits are obtained for energy barriers for the diffusion of
molecules which are much \emph{larger} than the barriers for desorption.
This means that the process of molecular diffusion on the surface can be
neglected in \emph{all} models.

Neglecting the two corresponding terms (diffusion of molecules to and
from sites of type $i$) in the model of molecules only,
\ref{ratefullmol}, we obtain the simple model
\begin{equation}
  \label{ratesimmol}
    \frac{\d n\s{i}}{\d t} = f(\rho\s{i}-n\s{i}) -W\s{i} n\s{i}.
\end{equation}
For irradiation with atoms instead, Eqs.~(\plainref{ratefull}), the
reduced model takes the form
\begin{subequations}
  \label{ratesim}
  \begin{align}
    \label{ratesima}
    \frac{\d n\p{H}}{\d t}
    &= f(1-n\p{H}) -W\p{H} n\p{H} -2D\p{H}n\p{H}^2,\\
    \label{ratesimb}
    \frac{\d n\ps{H_2}{i}}{\d t}
    &= \frac{\rho\s{i}-n\ps{H_2}{i}} {\sum_{j} (\rho\s{j}-n\ps{H_2}{j}) }
    D\p{H} n\p{H}^2
    -W\ps{H_2}{i} n\ps{H_2}{i}.
  \end{align}
\end{subequations}
As we show in the next section, these reduced models provide good fits
to the experimental data.  We use these models to obtain our results on
activation energies and to produce the fits seen in
Figures~\plainref{fig:HD-mol} to~\plainref{fig:D-at}.

\section{Analysis of Experimental Data}
\label{sec:analysis}

We now use the rate equations presented above to obtain the parameters
that describe the dynamics of atoms and molecules on the tholin surface.
The appropriate set of rate equations, i.e.,
Eqs.~(\plainref{ratefullmol}), (\plainref{ratefull}),
(\plainref{ratesimmol}) or~(\plainref{ratesim}), respectively, is
numerically integrated using a Runge-Kutta stepper.  The result of that
integration is a TPD curve that is then compared with the experimental
one.  The parameters are iteratively adjusted to obtain the best
agreement.

\subsection{Irradiation with Molecules}

The TPD curves of HD and D$_{2}$ molecules (\ref{fig:HD-mol,fig:D2-mol})
on tholin surfaces show a very broad distribution with two distinctive
peaks, or one large peak and a shoulder at higher temperatures.  The
large low-temperature peak is most accurately described as two
overlapping peaks, so that the entire curve is best described assuming
desorption of molecules from three different types of sites, each with
normally-distributed activation energies.

The fitting curves for HD and D$_2$ irradiation at surface temperatures
of $8$~K, $12$~K and $18$~K are presented in
\ref{fig:HD-mol,fig:D2-mol}.  The activation energies we find for
desorption of molecules are given in \ref{energytable}.  Recall that
diffusion of molecules has been found to be negligible in our fitting
procedure.
\begin{table}
  \centering
  \begin{tabular}{
      c@{\hspace{.8em}}
      c@{}
      c@{\hspace{.6em}}
      c@{}
      c@{\hspace{.6em}}
      c@{}
      c}
    \toprule
    & \multicolumn{6}{c}{Fraction of sites} \\
    & \multicolumn{2}{c}{$\rho_\mathrm{I}$}
    & \multicolumn{2}{c}{$\rho_\mathrm{II}$}
    & \multicolumn{2}{c}{$\rho_\mathrm{III}$} \\
    & \multicolumn{2}{c}{0.13}
    & \multicolumn{2}{c}{0.71}
    & \multicolumn{2}{c}{0.16} \\
    \midrule
    & \multicolumn{6}{c}{Mean$/$meV, standard deviation$/$meV} \\
    Molecule
    & $\edes\ps{mol}{\mathrm I}$
    & $\sigma_\mathrm{mol,I}$
    & $\edes\ps{mol}{\mathrm{II}}$
    & $\sigma_\mathrm{mol,II}$
    & $\edes\ps{mol}{\mathrm{III}}$
    & $\sigma_\mathrm{mol,III}$ \\
    \midrule
    HD & 30 & 3.3 & 40 & 4.3 & 55 & 4.5 \\
    D$_2$ & 30 & 3.3 & 42 & 4.3 & 58 & 4.5 \\ 
    \bottomrule
  \end{tabular}
  \caption{Surface Parameters Obtained by Fitting Rate Equation Solutions
    to TPD Curves}
  \label{energytable}
\end{table} 

The shift of the energy barriers from HD molecules to slightly higher
values for D$_2$ molecules can be understood in light of their different
atomic mass as discussed above.  The fact that the leading edge of the
$8$~K curves is not reproduced well by the fit is probably due to the
fact that the model does not include very shallow sites.
Though one can add more site types to account for these features, this
increases the computational cost and introduces additional fitting
parameters, reducing the conclusiveness of our results.  The essential
properties of the surface can be captured by the three energies found
here.

\subsection{Irradiation with Atoms}
\label{sec:atomsexp}

TPD curves of tholin surfaces irradiated with H and D atoms (H$+$D) or D
atoms (D$+$D) at several temperatures are presented in
\ref{fig:H-D-at,fig:D-at}, respectively.  In these experiments atoms are
deposited on the surface, where they diffuse and recombine.  The
resulting molecules stay on the surface---according to what is observed
experimentally---and desorb later during the TPD.  The only parameters
that need to be fitted to these experiments are the energy barriers for
diffusion and desorption of atoms, $E_{\mathrm H}^\mathrm{diff}$ and
$E_{\mathrm H}^\mathrm{des}$.

The experiments presented here were performed at three different
irradiation temperatures, $8$~K, $12$~K and $22$~K.  We observe that the
desorption peaks of the $8$~K experiments with atoms correspond to the
desorption peaks of the experiments with molecules.  Therefore, atoms
have to become sufficiently mobile to form molecules already during the
early stages of the TPD experiment.  An acceptable fit of the TPD data
is obtained only if the energy barrier for diffusion of atoms is at most
$E_{\mathrm H}^\mathrm{diff}=20$~meV.
The corresponding energy barrier for desorption of atoms is then found
as $E_{\mathrm H}^\mathrm{des}=30$~meV.

As seen in \ref{fig:H-D-at,fig:D-at}, the model correctly reproduces the
general tendency of the traces to shift to higher temperatures if the
irradiation temperature is increased.  For irradiation at $12$~K,
however, it lags behind the experimental data by $\approx 3$~K, and
the predicted peak position still closely resembles the one found for
irradiation with molecules.  When atoms are irradiated at a surface
temperature of $22$~K, the fit is very good again.

Our comments at the end of \ref{sec:results} might suggest that the
shift in TPD traces could be better reproduced by a model with several
types of sites for atoms (corresponding to the types for molecules).
Molecules produced on a certain type of site could then add specifically
to this type's population, and their distribution (affected by atomic
processes) might become more complex.  We checked that such a model
increases computational challenges, without improving the fits to
experimental data.

\section{Discussion}
\label{sec:applications}

\begin{figure}
  \centering
  \includegraphics{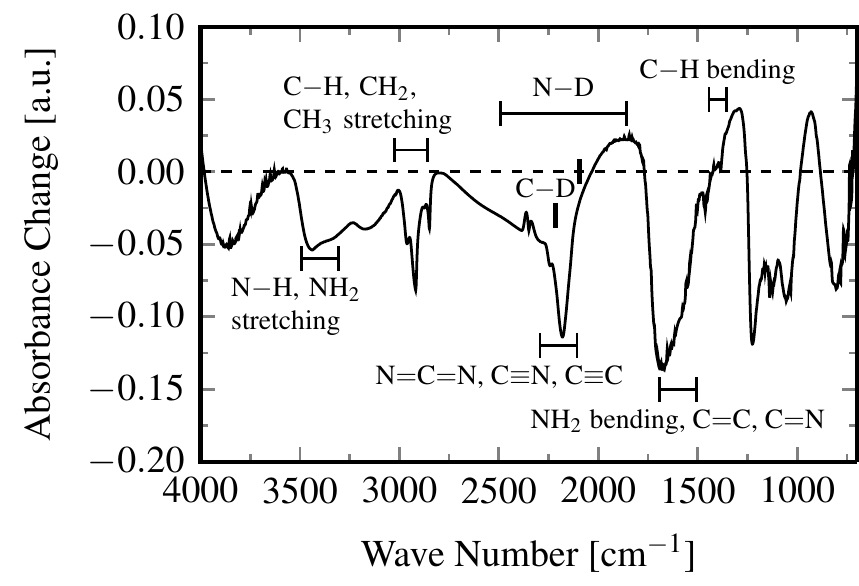}
  \caption{Change of infrared absorbance of the tholin sample after
    irradiation with H and D atoms for a cumulative dosage of $41$ and
    $64$ minutes, respectively.  Increased absorption is above the
    horizontal line.  Band assignments are indicated.  The
    sample has been kept under vacuum for the whole duration of the
    experiments.}
  \label{fig:FTIR-change}
\end{figure}

The formation of molecular hydrogen in space environments is of interest
to several research fields: to astrophysicists studying star formation,
since H$_2$ facilitates the cooling of a gravitationally collapsing
cloud by absorbing UV light and re-radiating it in the IR (where the
cloud is transparent); to astrochemists, because H$_2$ intervenes in
most schemes of formation of other molecules in space; and to planetary
scientists interested in the chemistry of atmospheres of bodies such as
Titan.  While the first experimental studies of H$_2$ formation were
geared towards the measurement of the efficiency of the
reaction~\cite{pirronello97a}, it was soon realized that one needed to
understand the elementary steps of atom and molecule adsorption,
diffusion on and desorption from heterogeneous surfaces.  In the case of
H interaction with tholins, experiments at higher temperatures imply
that the formation of HD is governed by the abstraction of surface-bound
H with D coming from the gas phase, and using IR spectroscopy it was
noticed that incoming D atoms saturate some of the carbon-carbon,
carbon-nitrogen, or nitrogen-nitrogen bonds of the
tholin~\cite{sekine08a}.  Our experiments of H and D interaction with
tholin surfaces are performed at a lower temperature and show that the
interaction is dominated by weak adsorption forces, and that the
production of HD is consistent with the LH reaction mechanism.
There are changes in the bonds of H and D with the surface, see
\ref{fig:FTIR-change}, due to partial saturation of C$-$H bonds with H
and D atoms, but, if there is abstraction of H, this process is
overwhelmed by the HD, H$_2$ and D$_2$ formation according to the
processes described above.  A detailed study of the changes of the IR
features and IR and TPD results with samples at higher temperature will
be presented elsewhere.  In the meantime, we can only speculate on how
our results and Sekine et al.'s~\cite{sekine08a} can be reconciled.
Aside from possible differences in the reactivity of samples, it is
possible that thermal energy atoms striking the surface at low
temperature experience the weak long-ranged physisorption interaction,
while other channels become available at higher surface temperature.  In
other words, H atoms might sample a precursor state.  Examples of this
behavior exist, such as H adsorption on Si~\cite{tok03}.

As an example of the application of our findings for the energy barriers
of hydrogen on tholins, we show the implied results for the efficiency
of H$_2$ formation on the surface of tholins in an environment
resembling Titan's atmosphere.  At a relevant height of about $700$~km,
we use a gas phase temperature of $T_\mathrm{gas} = 150$~K.  The average
thermal velocity of hydrogen atoms at this temperature is $v_\mathrm
H=1.8\times 10^5$~cm$/$s, and we take a density of $\rho_\mathrm H =
5.9\times10^7$~cm$^{-3}$ in the gas phase~\cite{lebonnois03}.  For the
density $s=5\times10^{14}$~sites$/$cm$^2$ assumed in \ref{sec:findflux}
we obtain a flux per site $f=\rho_\mathrm H v_\mathrm H/(4s)
=5.3\times10^{-3}$~ML$/$s.  The factor $1/4$ in the last relation
results from the ratio between the (geometrical) cross section and the
surface area of a sphere.

\begin{figure}
  \centering
  \includegraphics{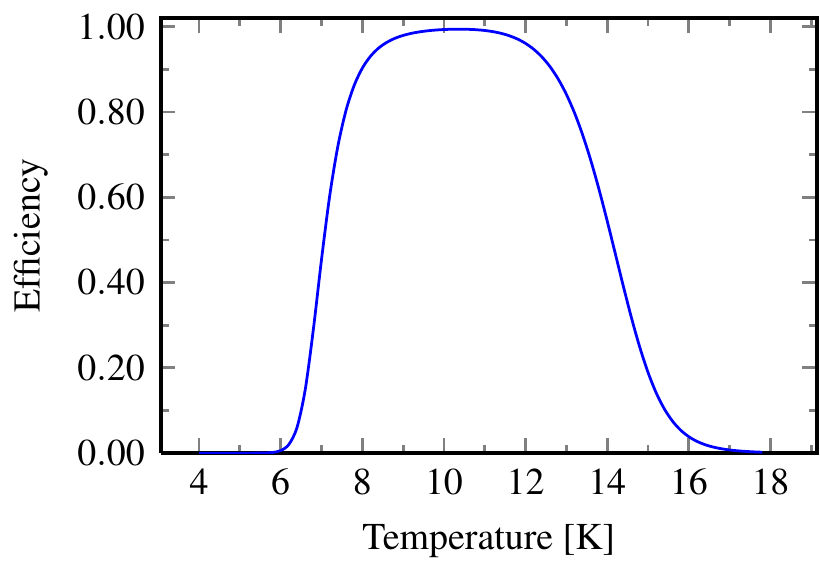}
  \caption{Efficiency of H$_2$ production on tholin as a function of the
    temperature for the surface parameters obtained in
    \ref{sec:atomsexp}.}
  \label{fig:efficiency}
\end{figure}

We calculate the efficiency $2D\p{H}n\p{H}^2/f$ using the analytic
solution of the steady-state rate equation, \ref{ratesima} with $\d
n\p{H}/\d t \equiv 0$, which provides accurate results for sufficiently
large grains.  \ref{fig:efficiency} shows that a high efficiency window
is found between $6$~K and $15$~K.  At temperatures below that window,
diffusion of atoms from one site to the other is so slow that they are
nearly immobile, and therefore the LH kinetics largely prevents sticking
of impinging atoms.  On the other hand at temperatures higher than
$\approx 15$~K, the residence time of H atoms is very short, they do not
encounter each other on the surface before they desorb, and the
efficiency drops as well.  For temperatures higher than this, other
mechanisms might take over in which one of the partners is held on the
surface by stronger adsorption forces.
Preliminary analysis of data taken at higher sample temperature
($>30$~K) shows that the HD formation rate is dramatically curtailed.
The quantification of the formation of HD at these higher sample
temperatures is ongoing.

\section{Summary}
\label{sec:summary}

We have studied the interaction of atomic and molecular hydrogen with
tholin surfaces, using TPD experiments at low temperatures.  Employing a
fine-grained rate equation model to fit TPD traces, we have obtained
energy barriers for the diffusion and desorption of both atomic as well
as molecular hydrogen.  The analysis shows that there are three types of
sites for molecules, and each type is associated with a distribution of
the energy barriers that can be fitted by a normal distribution.  In
contrast, there are no indications for a broad distribution of energy
barriers for the atoms, and the data can be fitted using a single
barrier.  All barriers are below $60$~meV, with the implication that the
interactions with the surface are only governed by weak adsorption
forces.
The temperature window of efficient formation of molecular hydrogen
depends on the diffusion and desorption barriers of the atoms, and not
on the interaction of the molecules with the surface.

\begin{acknowledgement}
  We are grateful to Prof.\ Mark Smith of the University of Arizona for
  providing the samples and to Dr.\ Imanaka (University of Arizona) for
  discussions.  We thank the US-Israel Binational Science Foundation for
  support.  G.V.\ was supported by NSF Grant AST-0507405.
\end{acknowledgement}


\begin{mcitethebibliography}{46}
\providecommand*{\natexlab}[1]{#1}
\providecommand*{\mciteSetBstSublistMode}[1]{}
\providecommand*{\mciteSetBstMaxWidthForm}[2]{}
\providecommand*{\mciteBstWouldAddEndPuncttrue}
  {\def\EndOfBibitem{\unskip.}}
\providecommand*{\mciteBstWouldAddEndPunctfalse}
  {\let\EndOfBibitem\relax}
\providecommand*{\mciteSetBstMidEndSepPunct}[3]{}
\providecommand*{\mciteSetBstSublistLabelBeginEnd}[3]{}
\providecommand*{\EndOfBibitem}{}
\mciteSetBstSublistMode{f}
\mciteSetBstMaxWidthForm{subitem}{(\alph{mcitesubitemcount})}
\mciteSetBstSublistLabelBeginEnd{\mcitemaxwidthsubitemform\space}
{\relax}{\relax}

\bibitem[Schlapbach and Z{\"u}ttel(2001)]{schlapbach01}
Schlapbach,~L.; Z{\"u}ttel,~A. \emph{Nature} \textbf{2001}, \emph{414},
  353--358\relax
\mciteBstWouldAddEndPuncttrue
\mciteSetBstMidEndSepPunct{\mcitedefaultmidpunct}
{\mcitedefaultendpunct}{\mcitedefaultseppunct}\relax
\EndOfBibitem
\bibitem[Combes and {Pineau des Forets}(2000)]{combes00}
\emph{Molecular Hydrogen in Space};
\newblock Combes,~F., {Pineau des Forets},~G., Eds.;
\newblock Cambridge University Press: Cambridge, 2000\relax
\mciteBstWouldAddEndPuncttrue
\mciteSetBstMidEndSepPunct{\mcitedefaultmidpunct}
{\mcitedefaultendpunct}{\mcitedefaultseppunct}\relax
\EndOfBibitem
\bibitem[Williams et~al.(2007)Williams, Brown, Price, Rawlings, and
  Viti]{williams07}
Williams,~D.~A.; Brown,~W.~A.; Price,~S.~D.; Rawlings,~J. M.~C.; Viti,~S.
  \emph{Astron. \& Geophys.} \textbf{2007}, \emph{48}, 1.25--1.34\relax
\mciteBstWouldAddEndPuncttrue
\mciteSetBstMidEndSepPunct{\mcitedefaultmidpunct}
{\mcitedefaultendpunct}{\mcitedefaultseppunct}\relax
\EndOfBibitem
\bibitem[Pirronello et~al.(1997)Pirronello, Liu, Shen, and
  Vidali]{pirronello97a}
Pirronello,~V.; Liu,~C.; Shen,~L.; Vidali,~G. \emph{Astrophys. J. Lett.}
  \textbf{1997}, \emph{475}, L69--72\relax
\mciteBstWouldAddEndPuncttrue
\mciteSetBstMidEndSepPunct{\mcitedefaultmidpunct}
{\mcitedefaultendpunct}{\mcitedefaultseppunct}\relax
\EndOfBibitem
\bibitem[Perets et~al.(2007)Perets, Lederhendler, Biham, Vidali, Li, Swords,
  Congiu, Roser, Manic{\`o}, Brucato, and Pirronello]{perets07}
Perets,~H.~B.; Lederhendler,~A.; Biham,~O.; Vidali,~G.; Li,~L.; Swords,~S.;
  Congiu,~E.; Roser,~J.; Manic{\`o},~G.; Brucato,~J.~R.; Pirronello,~V.
  \emph{Astrophys. J. Lett.} \textbf{2007}, \emph{661}, L163--166\relax
\mciteBstWouldAddEndPuncttrue
\mciteSetBstMidEndSepPunct{\mcitedefaultmidpunct}
{\mcitedefaultendpunct}{\mcitedefaultseppunct}\relax
\EndOfBibitem
\bibitem[Vidali et~al.(2009)Vidali, Li, Roser, and Badman]{vidali09}
Vidali,~G.; Li,~L.; Roser,~J.~E.; Badman,~R. \emph{Adv. Space. Res.}
  \textbf{2009}, \emph{43}, 1291--1298\relax
\mciteBstWouldAddEndPuncttrue
\mciteSetBstMidEndSepPunct{\mcitedefaultmidpunct}
{\mcitedefaultendpunct}{\mcitedefaultseppunct}\relax
\EndOfBibitem
\bibitem[Pirronello et~al.(1999)Pirronello, Liu, Roser, and
  Vidali]{pirronello99}
Pirronello,~V.; Liu,~C.; Roser,~J.~E.; Vidali,~G. \emph{Astron. Astrophys.}
  \textbf{1999}, \emph{344}, 681--686\relax
\mciteBstWouldAddEndPuncttrue
\mciteSetBstMidEndSepPunct{\mcitedefaultmidpunct}
{\mcitedefaultendpunct}{\mcitedefaultseppunct}\relax
\EndOfBibitem
\bibitem[Roser et~al.(2003)Roser, Swords, and Vidali]{roser03}
Roser,~J.~E.; Swords,~S.; Vidali,~G. \emph{Astrophys. J. Lett.} \textbf{2003},
  \emph{596}, L55--58\relax
\mciteBstWouldAddEndPuncttrue
\mciteSetBstMidEndSepPunct{\mcitedefaultmidpunct}
{\mcitedefaultendpunct}{\mcitedefaultseppunct}\relax
\EndOfBibitem
\bibitem[Hornek{\ae}r et~al.(2003)Hornek{\ae}r, Baurichter, Petrunin, Field,
  and Luntz]{hornekaer03}
Hornek{\ae}r,~L.; Baurichter,~A.; Petrunin,~V.~V.; Field,~D.; Luntz,~A.~C.
  \emph{Science} \textbf{2003}, \emph{302}, 1943--1946\relax
\mciteBstWouldAddEndPuncttrue
\mciteSetBstMidEndSepPunct{\mcitedefaultmidpunct}
{\mcitedefaultendpunct}{\mcitedefaultseppunct}\relax
\EndOfBibitem
\bibitem[Amiaud et~al.(2006)Amiaud, Fillion, Baouche, Dulieu, Momeni, and
  Lemaire]{amiaud06}
Amiaud,~L.; Fillion,~J.~H.; Baouche,~S.; Dulieu,~F.; Momeni,~A.; Lemaire,~J.~L.
  \emph{J. Chem. Phys.} \textbf{2006}, \emph{124}, 094702\relax
\mciteBstWouldAddEndPuncttrue
\mciteSetBstMidEndSepPunct{\mcitedefaultmidpunct}
{\mcitedefaultendpunct}{\mcitedefaultseppunct}\relax
\EndOfBibitem
\bibitem[Langmuir(1918)]{langmuir18}
Langmuir,~I. \emph{J. Am. Chem. Soc.} \textbf{1918}, \emph{40},
  1361--1403\relax
\mciteBstWouldAddEndPuncttrue
\mciteSetBstMidEndSepPunct{\mcitedefaultmidpunct}
{\mcitedefaultendpunct}{\mcitedefaultseppunct}\relax
\EndOfBibitem
\bibitem[Mennella(2008)]{mennella08}
Mennella,~V. \emph{Astrophys. J. Lett.} \textbf{2008}, \emph{684},
  L25--28\relax
\mciteBstWouldAddEndPuncttrue
\mciteSetBstMidEndSepPunct{\mcitedefaultmidpunct}
{\mcitedefaultendpunct}{\mcitedefaultseppunct}\relax
\EndOfBibitem
\bibitem[Sekine et~al.(2008)Sekine, Imanaka, Matsui, Khare, Bakes, McKay, and
  Sugita]{sekine08a}
Sekine,~Y.; Imanaka,~H.; Matsui,~T.; Khare,~B.~N.; Bakes,~E. L.~O.;
  McKay,~C.~P.; Sugita,~S. \emph{Icarus} \textbf{2008}, \emph{194},
  186--200\relax
\mciteBstWouldAddEndPuncttrue
\mciteSetBstMidEndSepPunct{\mcitedefaultmidpunct}
{\mcitedefaultendpunct}{\mcitedefaultseppunct}\relax
\EndOfBibitem
\bibitem[McKay et~al.(2001)McKay, Coustenis, Samuelson, Lemmon, Lorenz, Cabane,
  Rannou, and Drossart]{mckay01}
McKay,~C.~P.; Coustenis,~A.; Samuelson,~R.~E.; Lemmon,~M.~T.; Lorenz,~R.~D.;
  Cabane,~M.; Rannou,~P.; Drossart,~P. \emph{Planet. and Space Sci.}
  \textbf{2001}, \emph{49}, 79--99\relax
\mciteBstWouldAddEndPuncttrue
\mciteSetBstMidEndSepPunct{\mcitedefaultmidpunct}
{\mcitedefaultendpunct}{\mcitedefaultseppunct}\relax
\EndOfBibitem
\bibitem[Szopa et~al.(2006)Szopa, Cernogora, Boufendi, Correia, and
  Coll]{szopa06}
Szopa,~C.; Cernogora,~G.; Boufendi,~L.; Correia,~J.~J.; Coll,~P.
  \emph{Planetary and Space Science} \textbf{2006}, \emph{54}, 394--404\relax
\mciteBstWouldAddEndPuncttrue
\mciteSetBstMidEndSepPunct{\mcitedefaultmidpunct}
{\mcitedefaultendpunct}{\mcitedefaultseppunct}\relax
\EndOfBibitem
\bibitem[Khare et~al.(1984)Khare, Sagan, Thompson, Arakawa, Suits, Callcott,
  Williams, Shrader, Ogino, Willingham, and Nagy]{khare84}
Khare,~B.~N.; Sagan,~C.; Thompson,~W.~R.; Arakawa,~E.~T.; Suits,~F.;
  Callcott,~T.~A.; Williams,~M.~W.; Shrader,~S.; Ogino,~H.; Willingham,~T.~O.;
  Nagy,~B. \emph{Advances in Space Research} \textbf{1984}, \emph{4},
  59--68\relax
\mciteBstWouldAddEndPuncttrue
\mciteSetBstMidEndSepPunct{\mcitedefaultmidpunct}
{\mcitedefaultendpunct}{\mcitedefaultseppunct}\relax
\EndOfBibitem
\bibitem[Sagan and Khare(1979)]{sagan79}
Sagan,~S.; Khare,~B.~N. \emph{Nature} \textbf{1979}, \emph{277}, 102--107\relax
\mciteBstWouldAddEndPuncttrue
\mciteSetBstMidEndSepPunct{\mcitedefaultmidpunct}
{\mcitedefaultendpunct}{\mcitedefaultseppunct}\relax
\EndOfBibitem
\bibitem[Coll et~al.(1999)Coll, Coscia, Smith, Gazeau, Ram{\'\i}rez, Cernogora,
  Isra{\"e}l, and Raulin]{coll99}
Coll,~P.; Coscia,~D.; Smith,~N.; Gazeau,~M.-C.; Ram{\'\i}rez,~S.~I.;
  Cernogora,~G.; Isra{\"e}l,~G.; Raulin,~F. \emph{Planet. and Space Sci.}
  \textbf{1999}, \emph{47}, 1331--1340\relax
\mciteBstWouldAddEndPuncttrue
\mciteSetBstMidEndSepPunct{\mcitedefaultmidpunct}
{\mcitedefaultendpunct}{\mcitedefaultseppunct}\relax
\EndOfBibitem
\bibitem[Imanaka et~al.(2004)Imanaka, Khare, Elsila, Bakes, McKay, Cruikshank,
  Sugita, Matsui, and Zare]{imanaka04}
Imanaka,~H.; Khare,~B.~N.; Elsila,~J.~E.; Bakes,~E. L.~O.; McKay,~C.~P.;
  Cruikshank,~D.~P.; Sugita,~S.; Matsui,~T.; Zare,~R.~N. \emph{Icarus}
  \textbf{2004}, \emph{168}, 344--366\relax
\mciteBstWouldAddEndPuncttrue
\mciteSetBstMidEndSepPunct{\mcitedefaultmidpunct}
{\mcitedefaultendpunct}{\mcitedefaultseppunct}\relax
\EndOfBibitem
\bibitem[Quirico et~al.(2008)Quirico, Montagnac, Lees, McMillan, Szopa,
  Cernogora, Rouzaud, Simon, Bernard, Coll, Fray, Minard, Raulin, Reynard, and
  Schmitt]{quirico08}
Quirico,~E.; Montagnac,~G.; Lees,~V.; McMillan,~P.~F.; Szopa,~C.;
  Cernogora,~G.; Rouzaud,~J.-N.; Simon,~P.; Bernard,~J.-M.; Coll,~P.; Fray,~N.;
  Minard,~R.~D.; Raulin,~F.; Reynard,~B.; Schmitt,~B. \emph{Icarus}
  \textbf{2008}, \emph{198}, 218--231\relax
\mciteBstWouldAddEndPuncttrue
\mciteSetBstMidEndSepPunct{\mcitedefaultmidpunct}
{\mcitedefaultendpunct}{\mcitedefaultseppunct}\relax
\EndOfBibitem
\bibitem[Lebonnois et~al.(2003)Lebonnois, Bakes, and McKay]{lebonnois03}
Lebonnois,~S.; Bakes,~E. L.~O.; McKay,~C.~P. \emph{Icarus} \textbf{2003},
  \emph{161}, 474--485\relax
\mciteBstWouldAddEndPuncttrue
\mciteSetBstMidEndSepPunct{\mcitedefaultmidpunct}
{\mcitedefaultendpunct}{\mcitedefaultseppunct}\relax
\EndOfBibitem
\bibitem[Young et~al.(1984)Young, Allen, and Pinto]{young84}
Young,~Y.~L.; Allen,~M.; Pinto,~J.~P. \emph{Astrophys. J. Suppl.}
  \textbf{1984}, \emph{55}, 465\relax
\mciteBstWouldAddEndPuncttrue
\mciteSetBstMidEndSepPunct{\mcitedefaultmidpunct}
{\mcitedefaultendpunct}{\mcitedefaultseppunct}\relax
\EndOfBibitem
\bibitem[Duley and Williams(1984)]{duley84}
Duley,~W.~W.; Williams,~D.~A. \emph{Interstellar Chemistry};
\newblock Academic Press, 1984\relax
\mciteBstWouldAddEndPuncttrue
\mciteSetBstMidEndSepPunct{\mcitedefaultmidpunct}
{\mcitedefaultendpunct}{\mcitedefaultseppunct}\relax
\EndOfBibitem
\bibitem[Bakes et~al.(2003)Bakes, Lebonnois, Bauschlicher, and McKay]{bakes03}
Bakes,~E. L.~O.; Lebonnois,~S.; Bauschlicher,~C.~W.,~Jr.; McKay,~C.~P.
  \emph{Icarus} \textbf{2003}, \emph{161}, 468--473\relax
\mciteBstWouldAddEndPuncttrue
\mciteSetBstMidEndSepPunct{\mcitedefaultmidpunct}
{\mcitedefaultendpunct}{\mcitedefaultseppunct}\relax
\EndOfBibitem
\bibitem[Duley(1996)]{duley96}
Duley,~W.~W. \emph{Mon. Not. Roy. Astron. Soc.} \textbf{1996}, \emph{279},
  591--594\relax
\mciteBstWouldAddEndPuncttrue
\mciteSetBstMidEndSepPunct{\mcitedefaultmidpunct}
{\mcitedefaultendpunct}{\mcitedefaultseppunct}\relax
\EndOfBibitem
\bibitem[Harris and Kasemo(1981)]{harris81}
Harris,~J.; Kasemo,~B. \emph{Surf. Sci.} \textbf{1981}, \emph{105},
  L281--287\relax
\mciteBstWouldAddEndPuncttrue
\mciteSetBstMidEndSepPunct{\mcitedefaultmidpunct}
{\mcitedefaultendpunct}{\mcitedefaultseppunct}\relax
\EndOfBibitem
\bibitem[Rettner(1994)]{rettner94}
Rettner,~C.~T. \emph{J. Chem. Phys.} \textbf{1994}, \emph{101},
  1529--1546\relax
\mciteBstWouldAddEndPuncttrue
\mciteSetBstMidEndSepPunct{\mcitedefaultmidpunct}
{\mcitedefaultendpunct}{\mcitedefaultseppunct}\relax
\EndOfBibitem
\bibitem[Eilmsteiner et~al.(1996)Eilmsteiner, Walkner, and
  Winkler]{eilmsteiner96}
Eilmsteiner,~G.; Walkner,~W.; Winkler,~A. \emph{Surf. Sci.} \textbf{1996},
  \emph{352--354}, 263--267\relax
\mciteBstWouldAddEndPuncttrue
\mciteSetBstMidEndSepPunct{\mcitedefaultmidpunct}
{\mcitedefaultendpunct}{\mcitedefaultseppunct}\relax
\EndOfBibitem
\bibitem[Khan et~al.(2007)Khan, Takeo, Ueno, Inanaga, Yamauchi, Narita,
  Tsurumaki, and Namiki]{khan07}
Khan,~A.~R.; Takeo,~A.; Ueno,~S.; Inanaga,~S.; Yamauchi,~T.; Narita,~Y.;
  Tsurumaki,~H.; Namiki,~A. \emph{Surf. Sci.} \textbf{2007}, \emph{601},
  1635--1641\relax
\mciteBstWouldAddEndPuncttrue
\mciteSetBstMidEndSepPunct{\mcitedefaultmidpunct}
{\mcitedefaultendpunct}{\mcitedefaultseppunct}\relax
\EndOfBibitem
\bibitem[Zecho et~al.(2002)Zecho, G{\"u}ttler, Sha, Jackson, and
  K{\"u}ppers]{zecho02}
Zecho,~T.; G{\"u}ttler,~A.; Sha,~X.; Jackson,~B.; K{\"u}ppers,~J. \emph{J.
  Chem. Phys.} \textbf{2002}, \emph{117}, 8486\relax
\mciteBstWouldAddEndPuncttrue
\mciteSetBstMidEndSepPunct{\mcitedefaultmidpunct}
{\mcitedefaultendpunct}{\mcitedefaultseppunct}\relax
\EndOfBibitem
\bibitem[Vidali et~al.(2007)Vidali, Pirronello, Li, Roser, Manic{\'o}, Congiu,
  Mehl, Lederhendler, Perets, Brucato, and Biham]{vidali07}
Vidali,~G.; Pirronello,~V.; Li,~L.; Roser,~J.; Manic{\'o},~G.; Congiu,~E.;
  Mehl,~H.; Lederhendler,~A.; Perets,~H.~B.; Brucato,~J.~R.; Biham,~O. \emph{J.
  Phys. Chem. A} \textbf{2007}, \emph{111}, 12611--12619\relax
\mciteBstWouldAddEndPuncttrue
\mciteSetBstMidEndSepPunct{\mcitedefaultmidpunct}
{\mcitedefaultendpunct}{\mcitedefaultseppunct}\relax
\EndOfBibitem
\bibitem[Katz et~al.(1999)Katz, Furman, Biham, Pirronello, and Vidali]{katz99}
Katz,~N.; Furman,~I.; Biham,~O.; Pirronello,~V.; Vidali,~G. \emph{Astrophys.
  J.} \textbf{1999}, \emph{522}, 305--312\relax
\mciteBstWouldAddEndPuncttrue
\mciteSetBstMidEndSepPunct{\mcitedefaultmidpunct}
{\mcitedefaultendpunct}{\mcitedefaultseppunct}\relax
\EndOfBibitem
\bibitem[Manic{\`o} et~al.(2001)Manic{\`o}, Ragun{\`\i}, Pirronello, Roser, and
  Vidali]{manico01}
Manic{\`o},~G.; Ragun{\`\i},~G.; Pirronello,~V.; Roser,~J.~E.; Vidali,~G.
  \emph{Astrophys. J. Lett.} \textbf{2001}, \emph{548}, L253--256\relax
\mciteBstWouldAddEndPuncttrue
\mciteSetBstMidEndSepPunct{\mcitedefaultmidpunct}
{\mcitedefaultendpunct}{\mcitedefaultseppunct}\relax
\EndOfBibitem
\bibitem[Perets et~al.(2005)Perets, Biham, Manic{\'o}, Pirronello, Roser,
  Swords, and Vidali]{perets05}
Perets,~H.~B.; Biham,~O.; Manic{\'o},~G.; Pirronello,~V.; Roser,~J.;
  Swords,~S.; Vidali,~G. \emph{Astrophys. J.} \textbf{2005}, \emph{627},
  850--860\relax
\mciteBstWouldAddEndPuncttrue
\mciteSetBstMidEndSepPunct{\mcitedefaultmidpunct}
{\mcitedefaultendpunct}{\mcitedefaultseppunct}\relax
\EndOfBibitem
\bibitem[Vidali et~al.(2005)Vidali, Roser, Manic{\'o}, Pirronello, Perets, and
  Biham]{vidali05}
Vidali,~G.; Roser,~J.; Manic{\'o},~G.; Pirronello,~V.; Perets,~H.~B.; Biham,~O.
  \emph{J. Phys.: Conf. Ser.} \textbf{2005}, \emph{6}, 36--58\relax
\mciteBstWouldAddEndPuncttrue
\mciteSetBstMidEndSepPunct{\mcitedefaultmidpunct}
{\mcitedefaultendpunct}{\mcitedefaultseppunct}\relax
\EndOfBibitem
\bibitem[Neish et~al.(2010)Neish, Somogyi, and Smith]{neish10}
Neish,~C.~D.; Somogyi,~{\'A}.; Smith,~M.~A. \emph{Astrobiol.} \textbf{2010},
  \emph{10}, 337--347\relax
\mciteBstWouldAddEndPuncttrue
\mciteSetBstMidEndSepPunct{\mcitedefaultmidpunct}
{\mcitedefaultendpunct}{\mcitedefaultseppunct}\relax
\EndOfBibitem
\bibitem[Neish et~al.(2009)Neish, Somogyi, Lunine, and Smith]{neish09}
Neish,~C.~D.; Somogyi,~{\'A}.; Lunine,~J.~I.; Smith,~M.~A. \emph{Icarus}
  \textbf{2009}, \emph{201}, 412--421\relax
\mciteBstWouldAddEndPuncttrue
\mciteSetBstMidEndSepPunct{\mcitedefaultmidpunct}
{\mcitedefaultendpunct}{\mcitedefaultseppunct}\relax
\EndOfBibitem
\bibitem[Sarker et~al.(2003)Sarker, Somogyi, Lunine, and Smith]{sarker03}
Sarker,~N.; Somogyi,~A.; Lunine,~J.~I.; Smith,~M.~A. \emph{Astrobiol.}
  \textbf{2003}, \emph{3}, 719--726\relax
\mciteBstWouldAddEndPuncttrue
\mciteSetBstMidEndSepPunct{\mcitedefaultmidpunct}
{\mcitedefaultendpunct}{\mcitedefaultseppunct}\relax
\EndOfBibitem
\bibitem[Yates(1985)]{yates85}
Yates,~J.~T. In \emph{Methods of Experimental Physics, Solid State Physics:
  Surfaces};
\newblock Academic Press: New York, 1985;
\newblock Vol.~22, p 425\relax
\mciteBstWouldAddEndPuncttrue
\mciteSetBstMidEndSepPunct{\mcitedefaultmidpunct}
{\mcitedefaultendpunct}{\mcitedefaultseppunct}\relax
\EndOfBibitem
\bibitem[Barton et~al.(1978)Barton, Harrison, and Dollimore]{barton78}
Barton,~S.~S.; Harrison,~B.~H.; Dollimore,~J. \emph{J. Phys. Chem.}
  \textbf{1978}, \emph{82}, 290--294\relax
\mciteBstWouldAddEndPuncttrue
\mciteSetBstMidEndSepPunct{\mcitedefaultmidpunct}
{\mcitedefaultendpunct}{\mcitedefaultseppunct}\relax
\EndOfBibitem
\bibitem[Cazaux and Tielens(2004)]{cazaux04}
Cazaux,~S.; Tielens,~A. G. G.~M. \emph{Astrophys. J.} \textbf{2004},
  \emph{604}, 222--237\relax
\mciteBstWouldAddEndPuncttrue
\mciteSetBstMidEndSepPunct{\mcitedefaultmidpunct}
{\mcitedefaultendpunct}{\mcitedefaultseppunct}\relax
\EndOfBibitem
\bibitem[Barrie(2008)]{barrie08}
Barrie,~P.~J. \emph{Phys. Chem. Chem. Phys.} \textbf{2008}, \emph{10},
  1688--1696\relax
\mciteBstWouldAddEndPuncttrue
\mciteSetBstMidEndSepPunct{\mcitedefaultmidpunct}
{\mcitedefaultendpunct}{\mcitedefaultseppunct}\relax
\EndOfBibitem
\bibitem[Vidali and Li(2010)]{vidali10}
Vidali,~G.; Li,~L. \emph{J. Phys.: Condens. Matter} \textbf{2010}, \emph{22},
  304012\relax
\mciteBstWouldAddEndPuncttrue
\mciteSetBstMidEndSepPunct{\mcitedefaultmidpunct}
{\mcitedefaultendpunct}{\mcitedefaultseppunct}\relax
\EndOfBibitem
\bibitem[Ghio et~al.(1980)Ghio, Mattera, Salvo, Tommasini, and Valbusa]{ghio80}
Ghio,~E.; Mattera,~L.; Salvo,~C.; Tommasini,~F.; Valbusa,~U. \emph{J. Chem.
  Phys.} \textbf{1980}, \emph{73}, 556--561\relax
\mciteBstWouldAddEndPuncttrue
\mciteSetBstMidEndSepPunct{\mcitedefaultmidpunct}
{\mcitedefaultendpunct}{\mcitedefaultseppunct}\relax
\EndOfBibitem
\bibitem[Tok et~al.(2003)Tok, Engstrom, and Kang]{tok03}
Tok,~E.~S.; Engstrom,~J.~R.; Kang,~H.~C. \emph{J. Chem. Phys.} \textbf{2003},
  \emph{118}, 3294--3299\relax
\mciteBstWouldAddEndPuncttrue
\mciteSetBstMidEndSepPunct{\mcitedefaultmidpunct}
{\mcitedefaultendpunct}{\mcitedefaultseppunct}\relax
\EndOfBibitem
\end{mcitethebibliography}
\providecommand*{\mcitethebibliography}{\thebibliography}
\csname @ifundefined\endcsname{endmcitethebibliography}
{\let\endmcitethebibliography\endthebibliography}{}

\end{document}